# A Generalized Space-Frequency Index Modulation Scheme for Downlink MIMO Transmissions with Improved Diversity

Vasco Velez[1,2], João Pedro Pavia[1,2], Student Member, IEEE, Nuno Souto[1,2], Senior Member IEEE, Pedro Sebastião[1,2], Member IEEE, and Américo Correia[1,2] Senior Member, IEEE

[1]Department of Information Science and Technology, ISCTE-Instituto Universitário de Lisboa,1649-026 Lisboa, Portugal
[2]Instituto de Telecomunicações, 1049 - 001 Lisboa, Portugal
Corresponding author: Vasco Velez (e-mail: vasco_velez@iscte-iul.pt).

*Abstract*— Multidimensional Index Modulations (IM) are a novel alternative to conventional modulations which can bring considerable benefits for future wireless networks. Within this scope, in this paper we present a new scheme, named as Precoding-aided Transmitter side Generalized Space-Frequency Index Modulation (PT-GSFIM), where part of the information bits select the active antennas and subcarriers which then carry amplitude and phase modulated symbols. The proposed scheme is designed for multiuser multiple-input multiple-output (MU-MIMO) scenarios and incorporates a precoder which removes multiuser interference (MUI) at the receivers. Furthermore, the proposed PT-GSFIM also integrates signal space diversity (SSD) techniques for tackling the typical poor performance of uncoded orthogonal frequency division multiplexing (OFDM) based schemes. By combining complex rotation matrices (CRM) and subcarrier-level interleaving, PT-GSFIM can exploit the inherent diversity in frequency selective channels and improve the performance without additional power or bandwidth. To support reliable detection of the multidimensional PT-GSFIM we also propose three different detection algorithms which can provide different tradeoffs between performance and complexity. Simulation results shows that proposed PT-GSFIM scheme, can provide significant gains over conventional MU-MIMO and GSM schemes.

*Keywords — Index Modulation (IM), Precoding-aided Transmitter side Generalized Space-Frequency Index Modulation (PT-GSFIM), Multiple-Input Multiple-Output (MIMO), Orthogonal Frequency Division Multiplexing (OFDM), Multiuser-MIMO (MU-MIMO).*

I. INTRODUCTION

Due to the increased number of devices connected to the internet and to the amount of mobile traffic services growing up every day, new and more efficient solutions are needed to improve current wireless networks. Therefore, newer solutions are necessary to offer higher mobility communications with a higher spectral and energy efficiency (EE). This implies much higher transmission rates, with substantial lower latencies, enabling new types of applications such as: extended reality, internet of everything, autonomous connected vehicles and so on. Amongst the potential physical layer solutions for future wireless systems, index modulation (IM) schemes have been captivating a lot of research interest in the wireless community. Instead of using conventional multiple-input multiple-output (MIMO) with a large number of radio frequency (RF) chains and scaled power consumption, IM has attracted researchers as an alternative because of its interesting tradeoffs between number of data streams, performance and EE [1], [2]. These characteristics benefit both the receivers and transmitters enabling reduced complexity implementations [3], [4]. All these advantages make IM schemes potential candidates for future of wireless communications, namely for B5G and 6G networks [5].

IM techniques convey additional information bits contained implicitly in the index of the selected element (in the spatial, frequency, temporal/time or channel domain), such as the active antenna subset, as is done in generalized spatial modulations (GSM) [6] or the active subcarrier subset, as is done in IM aided orthogonal frequency division multiplexing (IM-OFDM) [7].

Spatial modulation (SM) is an IM scheme that is a simplification of MIMO with lower implementation and computational complexity, and it only requires a single RF chain to convey additional information [8]. In SM, only one antenna is active at any given time which often transmits a conventional amplitude and phase modulated (APM) symbol, such as phase shift keying (PSK) or quadrature amplitude modulation (QAM). Alternatively, information can be sent solely through the index of the active transmit antenna, as is done in space shift keying (SSK) modulation [9]. Comparatively, SSK can be seen as a low complexity implementation of SM since it does not require inter-antenna synchronization (IAS) and the complexity of the detection is smaller. Still, the spectral efficiency (SE) is also smaller than SM. GSM generalizes the concept of SM to multiple active antennas during transmission. Instead of activating a single transmit antenna, several antennas are activated simultaneously in order to transmit multiple *M*-ary modulated symbols [10], [11]. The additional information is conveyed implicitly through the respective indexes of the antenna-activation pattern. Even though SM schemes present some advantages over conventional MIMO, such as mitigating inter-channel interference (ICI) and better performance in terms of bit-error rate (BER) and EE, they also bring some drawbacks. In fact, the use of only one active antenna (or a reduced number in the case of GSM) limits the SE for the same number of available transmit antennas, also sacrificing some diversity gain [2], [11], [13].

Channel state information (CSI) is required in many MIMO schemes. It can be exploited at the transmitted side in order to reduce the complexity of receivers, and is referred to as CSIT, or at the receiver side, being referred to as CSIR, which is the most commonly used in SM/GSM schemes [6]. The use of CSIT combined with precoding can enable the use of SM, denoted as precoded SM (PSM), when there are strict complexity constraints at the receiver. In this case additional information bits are mapped to the index of the targeted receiver antenna during each transmission [12]. Generalized Precoding aided SM (GPSM) is an extension of PSM similar to GSM, which permits multiple receive antennas to be activated during each timeslot/transmission [13]. Quadrature SM (QSM) is another type of extension of SM, which divides the transmission into independent real and imaginary chunks that can be transmitted by different antennas simultaneously, with additional information bits being used for selecting the active antennas [14]. QSM is able to increase the SE of the system when compared to SM. Generalized Precoding aided QSM (GPQSM) generalizes the QSM concept to the receiver side, improving the SE over GPSM [15].

IM-OFDM, uses a similar approach to SM/GSM techniques but relies on the indices of existing subcarriers of a conventional OFDM transmission [16], [17]. In this case, only a subset of the subcarriers are activated depending on the index bits [18] and only these subcarriers convey $M$-ary signal modulated symbols [19], [20]. When index bits are triggered, they indicate which subcarriers inside a block of an OFDM frame should be activated during communication [4], [6]. In [21], the authors proposed a precoding technique, using IM on MIMO-OFDM systems, which tries to decrease the receiver complexity at the downlink. This is done by applying precoding to each subcarrier frequency so that the OFDM frames are received orthogonally without interference. A different approach is Layered Orthogonal Frequency Division Multiplexing with IM (L-OFDM-IM) which was proposed to increase SE in OFDM-IM scheme [22]. It splits the subcarriers into $n$ layers, with the index bits selecting active subcarriers that can overlap but are distinguishable through different carefully designed constellations. Recently, a lot of attention has been directed to schemes that involve the use of multiple signal constellations aided by IM [23] – [26]. The objective is to bring some improvements in SE and diversity to OFDM-IM schemes. In [23], it was introduced an OFDM-IM scheme named dual mode IM aided OFDM (DM-OFDM), that divides the subcarriers into two major groups, which are modulated with different constellations "modes", that correspond to two index subsets. Unlike an OFDM-IM system, where a subcarrier is switched on or off according to the index bits, DM-OFDM modulates all subcarriers which improves the SE [23]. In [24], generalized (G) DM-OFDM extends the previous idea to multiple constellations that are changeable according to the information bits of each OFDM subblock, with a minor loss in performance. Additionally, the authors in [25] proposed a novel multiple-mode (MM) OFDM-IM scheme where all subcarriers are activated with different types of PSK/QAM signal constellations. In order to improve the number of transmission patterns, the full permutations of the distinguishable modes are also used to convey additional information. This approach was generalized in [26] (G-MM-OFDM-IM) by allowing different subcarriers to carry signal constellations of different sizes while conveying the same number of IM bits. These schemes were shown to improve the SE over OFDM-IM and DM-OFDM at the cost of additional complexity.

Even though in this paper we focus on the use of IM in the space and frequency domain, there are other approaches proposed in the literature. For example, IM can also be applied in the time domain (TD), by using time slots which are activated according their indices, and can also be combined with space-time block codes (STBC) [27]. In [28], the authors presented a new type of IM-based uplink called index modulation–multiple access (IMMA), that uses a non-orthogonal multiple access (NOMA) method, operating with time slots similarly to transmission division multiple access (TDMA). In IMMA, each user can select his own time slot independently without the need of scheduling, but also, time slots can be shared with multiple users. Another different IM approach corresponds to channel domain index modulation (CD-IM), also known as media-based modulation (MBM), where information bits are transmitted via different channel realizations generated through available RF mirrors [29].

Regarding the research work that has been done in SM/GSM, besides single-user (SU) transmission, several multiuser (MU) schemes have been proposed for downlink and uplink scenarios. While the uplink can be addressed as a direct extension of SU GSM [30], the downlink requires a different approach. In the downlink, the users will typically have a limited number of antennas which can make them prone to MU interference (MUI). In this case one can resort to PSM to exploit the independence of multiple antennas and achieve higher SE and EE [31]. Resorting to precoding, the IM approach can be implemented at the transmitter side [34] or at the receiver side [31]. Receive SM (R-SM), consists in using some of information bits for selecting the receiver antennas, which will in fact receive the intended signals [31] – [33]. A specific precoder or special preprocessing stages must be adopted to achieve the desired antenna activation pattern at the receiver. One drawback of this approach concerns the fact that the receiver antennas must remain active all the time, even though only a portion of them receive symbols [32]. In [34], the authors adopted a different form of precoded space domain IM, which tries to improve SE by using a virtual GSM applied at the transmitter side through the use of precoding. In this case, virtual active antennas transmit $M$-ary modulated symbols using a precoder in order to mitigate MUI and allow SU GSM detection.

Besides the different versions that IM can assume, which depend on the specific dimension that is exploited, several authors have recently started to propose multidimensional schemes that are based on combinations of multiple one-dimensional IM [4]. For example, in generalized space-frequency IM (GSFIM) [35], both SM and OFDM-IM schemes are combined to make use of the frequency domain and spatial domain simultaneously. In this case, index bits are used for selecting the active antennas and subcarriers at each transmission interval. The authors in [35] explored this approach showing good results both in achievable rates and BER when compared with conventional schemes. However, only SU scenarios were considered. One problem with OFDM, and consequently with OFDM-IM and GSFIM is that it does not exploit the frequency diversity which can degrade the performance when some subcarriers experience deep fades. In this case, low-rate channel codes can be adopted but this reduces the system's SE. As an alternative, in [36] the authors proposed the use of subcarrier interleaving for improving the performance of OFDM-IM.

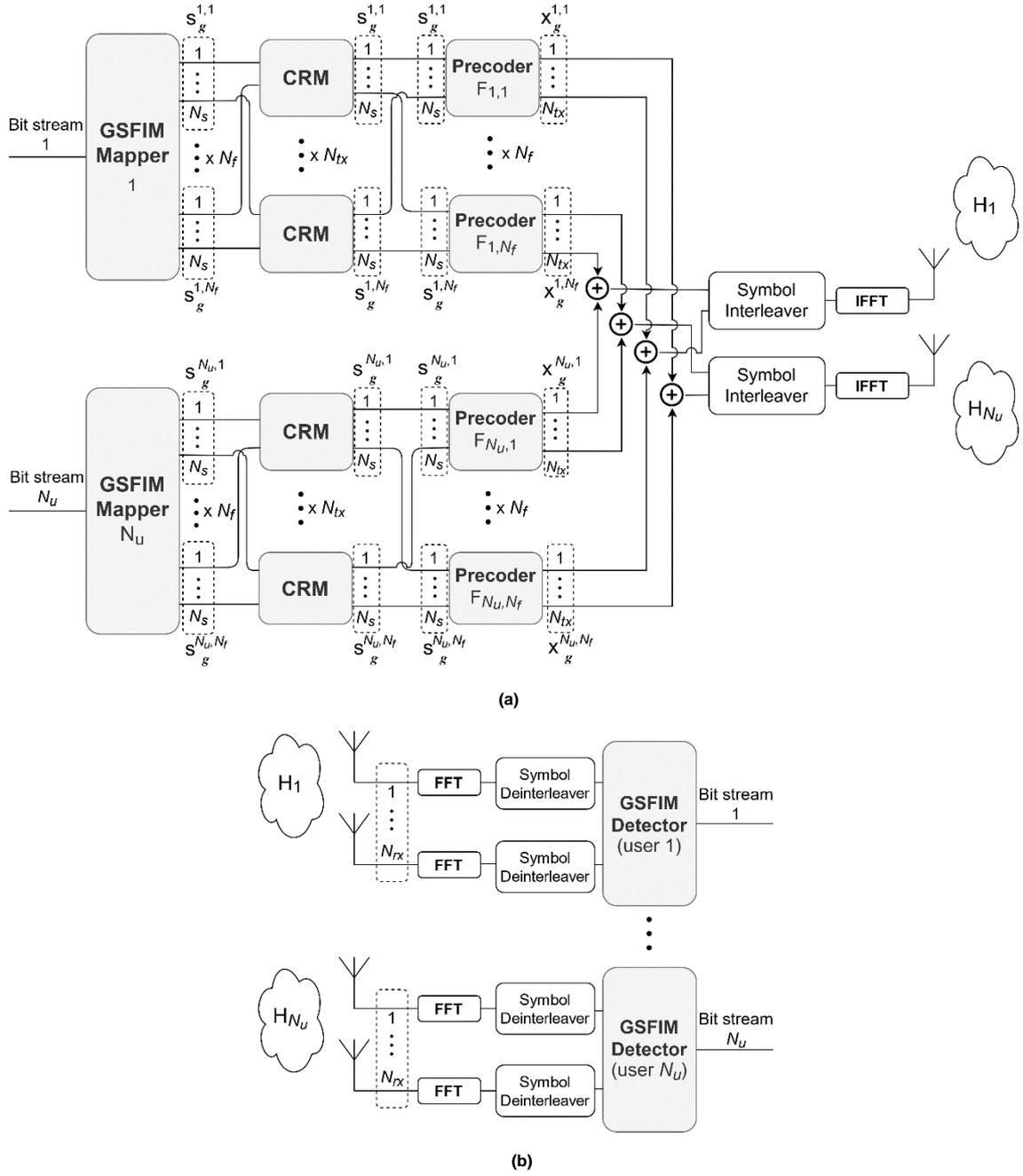

Fig 1. System Model: PT-GSFIM transmitter (a) and receiver (b).

Motivated by the work above, in this paper we assume an OFDM MU downlink transmission as the baseline and propose a precoding-aided, transmitter side multi-dimensional IM scheme which we refer to as PT-GSFIM. In PT-GSFIM, part of the information bits are used to select resources in the spatial and frequency domain whereas the remainder of the bits APM symbols. To help circumvent the poor performance problem of uncoded GSFIM we adopt the concept of signal space diversity (SSD), which was originally presented in [37], into the design of the signal. The use of SSD enables us to associate the bits over several subcarriers and benefit from the diversity effects inherent to a frequency selective channel, while keeping the mapping/de-mapping process simple. In this case, we integrate SSD into PT-GSFIM using complex rotation matrices (CRM), which were proposed in [38] to achieve time-diversity within the context of downlink transmission in a cellular system with multiple transmit antennas. At the transmitter side we consider that a precoder is used for removing MUI. Regarding the receiver, we propose several types of detectors with different tradeoffs in terms of complexity and performance. The main contributions of this paper can be summarized as follows:

- A frequency and space domain IM scheme that we refer to as PT-GSFIM, is designed for the downlink of MU-MIMO systems. To simplify the implementation, PT-GSFIM independently encodes part of the information bits onto spatial indexes (selecting active virtual antennas) and part onto frequency indexes (active subcarriers). The remaining information bits are mapped onto *M*-ary modulated symbols. To handle MU downlink scenarios, PT-GSFIM incorporates a precoder applied to each subcarrier but to the whole set of transmit antennas (active and inactive) which keeps the local structure of a conventional transmitted GSM signal intact. This allows the removal of MUI and transforms a MU communication into multiple independent SU communications.

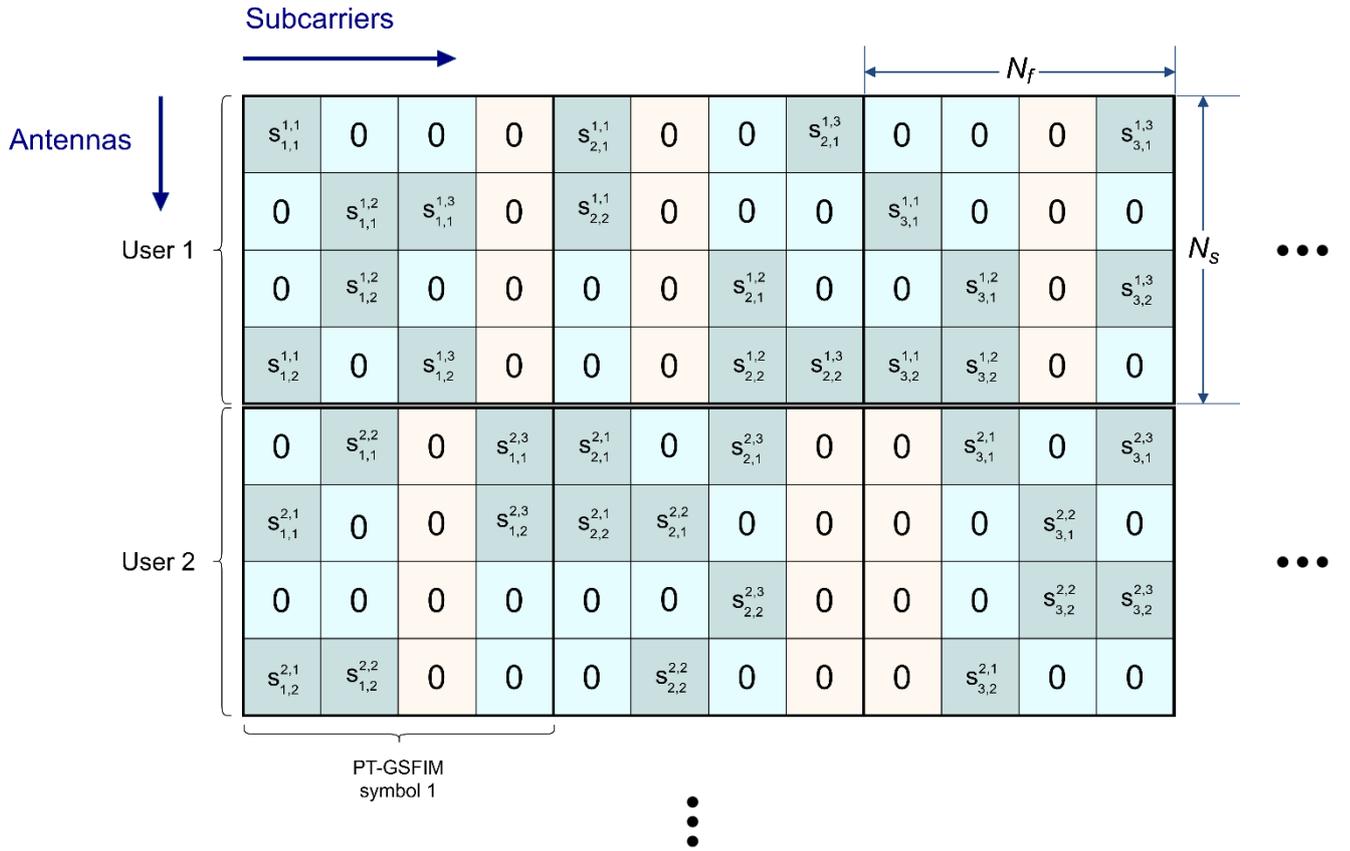

Fig 2. Example of the structure of each PT-GSFIM symbol in the space-frequency domain with $N_f=4$, $N_{af}=3$, $N_s=4$, $N_a=2$.

- To avoid the problem of having active subcarriers experiencing deep fades at the same time that some inactive subcarriers experience stronger channels, the proposed PT-GSFIM integrates SSD, implemented using CRM matrices applied over each sub-block. Combined with subcarrier level interleaving, this approach can exploit the inherent diversity in frequency selective channels whilst keeping the structure of the PT-GSFIM signal intact from the point of view of the receivers. Improved performance can thus be obtained without any additional power or bandwidth.
- To accomplish reliable detection of the multidimensional PT-GSFIM signal at the receiver's side, three different algorithms are presented. The proposed approaches are generalized versions of three known GSM algorithms that were modified in order to cope with the special structure of PT-GSFIM signals with CRM, which must consider the joint combination of active antennas and subcarriers. It is shown that the three derived algorithms provide different tradeoffs between performance and complexity.

The paper is organized as follows: Section II presents the system model which is followed by the PT-GSFIM transmitter structure formulation in section III. The three different algorithms proposed for signal detection are described in section IV alongside their respective complexities. Performance results are then presented in section V. Finally, the conclusions are outlined in section VI.

*Notation:* Matrices and vectors are denoted by uppercase and lowercase boldface letters, respectively. $(\cdot)^T$ and $(\cdot)^H$ denote the transpose and conjugate transpose of a matrix/vector, $\|\cdot\|_p$ is the $\ell_p$-norm of a vector, $\|\cdot\|_0$ is its cardinality, $\text{supp}(\mathbf{x})$ returns the support of $\mathbf{x}$, $\lfloor \cdot \rfloor$ is the floor function and $\mathbf{I}_n$ is the $n \times n$ identity matrix.

## II. SYSTEM MODEL

Let us consider the downlink of an OFDM-based MU-MIMO system where a BS equipped with $N_{tx}$ antennas transmits to $N_u$ users, each with $N_{rx}$ antennas, as shown in Fig. 1.

We consider also the adoption of an OFDM scheme, where part of the information is mapped onto conventional *M*-sized APM symbols and the other part is encoded both on spatial and frequency indexes. While the encoding onto the spatial indexes follows the same GSM approach from [39], the mapping onto the frequency indexes assumes the grouping of the subcarriers into $N_f$-sized subblocks inside which only $N_{af}$ subcarriers are active for each user. We assume a simple mapping of the bits to the transmitted block matrix, where the encoding onto spatial indexes is accomplished independently from the encoding to the frequency indexes, using simple look-up tables (LUTs): one for the spatial resources and one for the frequency resources in each subblock. While this independent mapping sacrifices some SE (the number of available space-frequency blocks that can be indexed is reduced) it makes the mapping/de-mapping process easier as well as the detection, as we will show further ahead. It is important to note also that the approach based on a pair of LUTs is adequate when the number of active antenna combinations and active subcarrier combinations is small, as is the case of the scenarios considered in this paper. If the number of combinations is large, then a combination strategy can be employed as described in [40] and [20]. Fig. 2 illustrates an example of the structure adopted for each PT-GSFIM symbol for the case of frequency domain subblocks with $N_f=4$ subcarriers each, where $N_{af}=3$ of them are active subcarriers, thus encoding frequency domain index bits. These active subcarriers convey GSM symbols ([39]) which encode both spatial bits and conventional modulated symbols, and are thus used as the base element for constructing the whole PT-GSFIM symbol. In this example each user's GSM symbol has a size of $N_s=4$ with $N_a=2$ active (nonzero) positions.

According to the figure, the $g^{th}$ PT-GSFIM symbol, $\mathbf{s}_g^u \in \mathbb{C}^{N_s N_f \times 1}$, for the $u^{th}$ user can be written as $\mathbf{s}_g^u = \left[ \left( \mathbf{s}_g^{u,1} \right)^T \ldots \mathbf{0} \ldots \left( \mathbf{s}_g^{u,N_{af}} \right)^T \right]^T$, which corresponds to the concatenation of ($N_f$-$N_{af}$) length-$N_s$ vectors of zeros with $N_{af}$ GSM symbol vectors $\mathbf{s}_g^{u,i} \in \mathbb{C}^{N_s \times 1}$ that are defined as $\mathbf{s}_g^{u,i} = \begin{bmatrix} 0 & s_{g,1}^{u,i} & 0 & \cdots & 0 & s_{g,N_a}^{u,i} & 0 \end{bmatrix}^T$, where $i=1,...,N_{af}$, $u=1,...,N_u$ and $g=1,...,N_{GSFIM}$. The symbols $s_{g,j}^{u,i}$, $j=1,...,N_a$ are selected from an APM constellation. Assuming a MU downlink scenario where precoding is applied, we can write the transmitted signal as

$$\mathbf{x}_g = \sum_{u=1}^{N_u} \mathbf{F}_u \mathbf{s}_g^u = \mathbf{F} \mathbf{s}_g, \quad (1)$$

where $\mathbf{s}_g = \left[ \left( \mathbf{s}_g^1 \right)^T \ldots \left( \mathbf{s}_g^{N_u} \right)^T \right]^T$. Matrix $\mathbf{F} = \begin{bmatrix} \mathbf{F}_1 \ldots \mathbf{F}_{N_u} \end{bmatrix}$, with $\mathbf{F}_u \in \mathbb{C}^{N_{tx} N_f \times N_s N_f}$, denotes the precoder matrix. After the precoders (and interleavers, which are explained further ahead), we consider that the frequency domain symbols are grouped into $N$-sized blocks and converted to the time domain by the Inverse Fast Fourier Transform (IFFT) blocks. A cyclic prefix (CP) with length larger than the delay spread of the channel is then added to each block before transmission. Note that each of the described PT-GSFIM symbols carries a total of bits per user given by

$$N_{bits} = N_{af} \left( \left\lfloor \log_2 \binom{N_s}{N_a} \right\rfloor + N_a \log_2 M \right) + \left\lfloor \log_2 \binom{N_f}{N_{af}} \right\rfloor. \quad (2)$$

To illustrate, if 64-QAM is employed in the case of Fig. 2, a total of 47 bits are mapped to a symbol, i.e., an average of 11.75 bits per subcarrier are being transmitted to each user.

## III. TRANSMITTER STRUCTURE

### A. Precoder Design

Assuming that the CP has been dropped and the time domain samples have been converted to the frequency domain through an $N$-point Fast Fourier Transform (FFT), the $g^{th}$ PT-GSFIM symbol received by the $u^{th}$ user can be written as

$$\mathbf{y}_g^u = \mathbf{H}_g^u \mathbf{x}_g + \mathbf{n}_g^u = \mathbf{H}_g^u \mathbf{F}_u \mathbf{s}_g^u + \mathbf{H}_g^u \sum_{\substack{j=1 \\ j \neq u}}^{N_u} \mathbf{F}_j \mathbf{s}_g^j + \mathbf{n}_g^u, \quad (3)$$

where $\mathbf{y}_g^u \in \mathbb{C}^{N_{rx} N_f \times 1}$, $\mathbf{n}_g^u \in \mathbb{C}^{N_{rx} N_f \times 1}$ represents noise and $\mathbf{H}_g^u \in \mathbb{C}^{N_{rx} N_f \times N_{tx} N_f}$ is the frequency domain channel matrix between the BS and user $u$. Since we are considering an OFDM system, the channel matrix follows a block diagonal structure, i.e,

$$\mathbf{H}_g^u = \text{blkdiag} \left\{ \mathbf{H}_g^{u,1}, ..., \mathbf{H}_g^{u,f}, ..., \mathbf{H}_g^{u,N_f} \right\} \quad (4)$$

with $\mathbf{H}_g^{u,f} \in \mathbb{C}^{N_{rx} \times N_{tx}}$, $f=1,...,N_f$. In the right side of expression (3), the second term represents interference between users. Each of the precoder matrices can be designed in order to eliminate all this MUI at all the receivers following the block diagonalization approach described in [43]. However, since many positions of the PT-GSFIM symbol, $\mathbf{s}_g^u$, are empty, with part of the information encoded in the equivalent channel impulse responses, we do not apply any of the power loading optimization approaches from [43] (even though power control can still be applied between users). Due to the block diagonal structure of $\mathbf{H}_g^u$, the signals transmitted on each subcarrier only generate interference on that subcarrier.

This means that the precoder matrices $\mathbf{F}_u$ can be designed independently for each subcarrier and follow a block diagonal structure, namely

$$\mathbf{F}_u = \text{blkdiag} \left\{ \mathbf{F}_{u,1}, ..., \mathbf{F}_{u,N_f} \right\}, \quad (5)$$

with $\mathbf{F}_{u,f} \in \mathbb{C}^{N_{tx} \times N_s}$, $f=1,...,N_f$. To cancel all MUI, each of the component precoder matrices $\mathbf{F}_{u,f}$ is designed so as to enforce $\mathbf{H}_g^{v,f} \mathbf{F}_{u,f} = \mathbf{0}$ for all $v \neq u$. Let us define the matrix that concatenates all the channel matrices between the BS and all users except user $k$, in subcarrier $f$ as

$$\tilde{\mathbf{H}}_g^{u,f} = \left[ \left( \mathbf{H}_g^{1,f} \right)^T \ldots \left( \mathbf{H}_g^{u-1,f} \right)^T \left( \mathbf{H}_g^{u+1,f} \right)^T \ldots \left( \mathbf{H}_g^{N_u,f} \right)^T \right]^T. \quad (6)$$

This matrix models the propagation of the signal targeted to user $u$, when it arrives at all the other receivers where it will constitute interference, therefore it can be referred to as the interference channel matrix. To avoid the generated interference one can design the respective precoder matrix $\mathbf{F}_{u,f}$ using an orthonormal basis of the null space of $\tilde{\mathbf{H}}_g^{u,f}$. First, we compute the respective singular value decomposition (SVD) given by

$$\tilde{\mathbf{H}}_g^{u,f} = \tilde{\mathbf{U}}_{u,f} \tilde{\mathbf{\Lambda}}_{u,f} \left[ \tilde{\mathbf{V}}_{u,f}^{(1)} \ \tilde{\mathbf{V}}_{u,f}^{(0)} \right]^H, \quad (7)$$

where $\tilde{\mathbf{U}}_{u,f}$ is the matrix with the left-singular vectors, $\tilde{\mathbf{\Lambda}}_{u,f}$ is a rectangular diagonal matrix with the decreasing nonzero singular values, $\tilde{\mathbf{V}}_{u,f}^{(1)}$ contains the right singular vectors corresponding to the nonzero singular values and $\tilde{\mathbf{V}}_{u,f}^{(0)}$ contains the remainder right singular vectors. Since the columns of $\tilde{\mathbf{V}}_{u,f}^{(0)}$ span the null space of $\tilde{\mathbf{H}}_g^{u,f}$, to guarantee zero MUI each precoder matrix can be set as $\mathbf{F}_{u,f} = \tilde{\mathbf{V}}_{u,f}^{(0)}[:,1:N_s]$. The resulting signal at each receiver then reduces to

$$\mathbf{y}_g^u = \hat{\mathbf{H}}_g^u \mathbf{s}_g^u + \mathbf{n}_g^u, \quad (8)$$

where $\hat{\mathbf{H}}_g^u = \mathbf{H}_g^u \mathbf{F}_u$ is the equivalent SU channel and no MUI exists. Bearing in mind that both $\mathbf{H}_g^u$ and $\mathbf{F}_u$ have a block diagonal structure, $\hat{\mathbf{H}}_g^u$ will also be block diagonal. According to the presented system model, only a few positions of $\mathbf{s}_g^u$ will tend to be nonzero. Therefore, $\mathbf{s}_g^u$ is a sparse vector and this sparsity can be exploited for reducing the transmitted power. To accomplish this, we can apply the same approach that we discussed in [34] for virtual GSM transmissions, where an alternative signal is generated which minimizes the transmitted power while ensuring that the signals arriving at the receivers are identical to the original ones.

Although in this paper we are assuming fully-digital precoders, it is possible to reduce the implementation complexity of the proposed solution by adopting an hybrid precoder design where the signal processing is split into two separate parts: a reduced digital one and an analog part which is typically supported on analog phase shifters. The hybrid design can be simply achieved through direct approximation of the fully-digital precoder matrices using the product of smaller digital precoder matrices and an analog precoder matrix (the same for all subcarriers), as described in [41] and [42]. As also explained in [42], an additional cancellation step should be included in this approximation-based approach in order to remove residual inter-user interference.

### B. Complex Rotation Matrices

According to the signal model (8) where $\hat{\mathbf{H}}_g^u$ is block diagonal, each individual GSM symbol vector composing a PT-GSFIM is subject to the effect of the channel on a single subcarrier. In order to exploit the inherent diversity of frequency selective channels that are typical in mobile propagation environments, we can apply SSD techniques. SSD was originally proposed in [37] and can be used in an OFDM system so as to improve its performance without requiring additional power or bandwidth. A simple way to implement SSD is to resort to CRM so as to associate each GSM symbol that makes up the main PT-GSFIM symbol to different subcarriers instead of only one. The process of applying CRM consists in working with a rotated super-symbol. In this case we apply the spreading over the frequency direction only, i.e., all the different spatial components are subject to the same rotation. The rotated length-$N_s N_f$ super-symbol for each user $u$ can be expressed as

$$\bar{\mathbf{s}}_g^u = \left(\mathbf{A}_{N_f} \otimes \mathbf{I}_{N_s}\right) \cdot \mathbf{s}_g^u, \quad (9)$$

and the transmitted signal becomes (1)

$$\mathbf{x}_g = \sum_{u=1}^{N_u} \mathbf{F}_u \left(\mathbf{A}_{N_f} \otimes \mathbf{I}_{N_s}\right) \mathbf{s}_g^u = \mathbf{F}\left(\mathbf{I}_{N_u} \otimes \left(\mathbf{A}_{N_f} \otimes \mathbf{I}_{N_s}\right)\right) \mathbf{s}_g. \quad (10)$$

Matrix $\mathbf{A}_{N_f} \in \mathbb{C}^{N_f \times N_f}$ can be selected from the family of orthonormal complex matrices (OCRM), which are defined as

$$\mathbf{A}_{M_{CRM}} = \begin{cases} \begin{bmatrix} e^{j\varphi} & je^{-j\varphi} \\ je^{j\varphi} & e^{-j\varphi} \end{bmatrix} / |\mathbf{A}_2|^{1/2}, M_{CRM} = 2 \\ |\mathbf{A}_2| = \det(A_2) = 2 \\ \begin{bmatrix} \mathbf{A}_{M_{CRM}/2} & \mathbf{A}_{M_{CRM}/2} \\ \mathbf{A}_{M_{CRM}/2} & -\mathbf{A}_{M_{CRM}/2} \end{bmatrix} / |\mathbf{A}_{M_{CRM}}|^{1/M_{CRM}}, M_{CRM} > 2 \end{cases} \quad (11)$$

where $M_{CRM} = 2^n$ ($n \geq 1$), $|\mathbf{A}_{M_{CRM}}| = \det(\mathbf{A}_{M_{CRM}})$ and $\varphi$ being the rotation angle [38]. To better reap the benefits of frequency diversity, a symbol interleaver (subcarrier-wise) should be employed so that the effective group of subcarriers allocated to each PT-GSFIM symbol is spread far apart within the overall bandwidth.

### IV. SIGNAL DETECTION

#### A. Proposed Algorithms

According to the transmitted structure described in the previous section, where a precoder removes all the MUI, the receiver only needs to implement SU detection. If no CRM is employed, the block diagonal structure of $\hat{\mathbf{H}}_g^u$ in (8) allows us to simplify the detection process. In this case the techniques proposed for GSM, such as the ordered block minimum mean-squared error (OB-MMSE) detector from [44], the multipath matching pursuit with slicing (sMMP) from [45] or the alternating direction method of multipliers (ADMM) based detector from [39], can be directly applied to PT-GSFIM by simply preceding those algorithms by an active subcarrier detection step. This initial detection can be accomplished by working with matrix $\mathbf{Y} = vec^{-1}_{N_{rx} \times N_f}\left(\mathbf{y}_g^u\right)$, where $vec^{-1}_{m \times n}(\ )$ denotes the inverse of the vectorization operator $vec(\ )$, which is defined as $vec^{-1}_{m \times n}: \mathbb{C}^{mn \times 1} \to \mathbb{C}^{m \times n}$ such that $vec^{-1}_{m \times n}(vec(\mathbf{X})) = \mathbf{X}$ for any $\mathbf{X} \in \mathbb{C}^{m \times n}$. The indexes of the $N_{af}$ columns of $\mathbf{Y}$ with larger Euclidean norm and which also match a valid active subcarrier combination are selected as those which will be processed. Each of these columns, together with the corresponding channel matrix $\mathbf{H}_g^{u,f}$ are then used as inputs for the GSM detection algorithms mentioned previously, which will estimate the active antennas and APM symbols for that individual subcarrier. In the case of CRM being employed, a different approach must be adopted since each GSM symbol is spread over several subcarriers. In this case the received signal can be written as

$$\begin{aligned}\mathbf{y}_g^u &= \hat{\mathbf{H}}_g^u \bar{\mathbf{s}}_g^u + \mathbf{n}_g^u \\ &= \breve{\mathbf{H}}_g^u \mathbf{s}_g^u + \mathbf{n}_g^u,\end{aligned} \quad (12)$$

TABLE I
ALGORITHM 1: OB-MMSE-BASED GSFIM DETECTOR

1: **Input:** $\mathbf{y}_g^u$, $\breve{\mathbf{H}}_g^u$, $\mathbb{I}$, $V_{th} = 2N_{rx}\sigma^2$.
2: $z_j = \dfrac{1}{\left(\breve{\mathbf{H}}_g^u[:,j]\right)^H \breve{\mathbf{H}}_g^u[:,j]} \left(\breve{\mathbf{H}}_g^u[:,j]\right)^H \mathbf{y}_g^u$, $j = 1,...,N_f N_s$
3: $\mathbf{w} = \left[w_1, w_2, ..., w_{N_{comb}}\right]^T$,
$w_i = \sum_{k \in I_i}^{N_{af}N_a} z_k^2$, $I_i \in \mathbb{I}$, $i = 1,2,...,N_{comb}$
4: $\left[k_1, k_2, ..., k_{N_{comb}}\right] = \arg\text{sort}(\mathbf{w})$.
5: $j = 1$.
6: **While** $j \leq N_{comb}$ **do**
7: $\tilde{\mathbf{s}}_g^u[\overline{I}_j] \leftarrow 0$
$\tilde{\mathbf{s}}_g^u[I_j] \leftarrow \Pi_{\mathcal{A}^{N_{af}N_a}}\left(\left(\left(\breve{\mathbf{H}}_g^u[:,I_j]\right)^H \breve{\mathbf{H}}_g^u[:,I_j] + 2\sigma^2 \mathbf{I}_{N_{af}N_s}\right)^{-}\right.$
$\left.\times \left(\breve{\mathbf{H}}_g^u[:,I_j]\right)^H \mathbf{y}_g^u\right)$
$d_j = \left\|\mathbf{y}_g^u - \breve{\mathbf{H}}_g^u[:,I_j]\cdot\tilde{\mathbf{s}}_g^u[I_j]\right\|_F^2$.
8: **If** $d_j < V_{th}$ **then**
9: $\hat{\mathbf{s}}_g^u[\overline{I}_j] \leftarrow 0$, $\hat{\mathbf{s}}_g^u[I_j] \leftarrow \tilde{\mathbf{s}}_g^u[I_j]$, break
10: **else**
11: $j = j+1$;
12: **end if**
13: **end while**
14: **If** $j > N_{comb}$ **then**
15: $m = \arg\min_{j \in \{1,...,N_{comb}\}} d_j$,
$\hat{\mathbf{s}}_g^u[\overline{I}_m] \leftarrow 0$, $\hat{\mathbf{s}}_g^u[I_m] \leftarrow \tilde{\mathbf{s}}_g^u[I_m]$.
16: **end if**
17: **Output:** $\hat{\mathbf{s}}_g^u$.

TABLE II
ALGORITHM 2: MULTIBLOCK SMMP DETECTOR

1: **Input:** $\mathbf{y}_g^u$, $\breve{\mathbf{H}}_g^u$, $L$;
2: **Initial:** $k = 0, r_1^{(0)} = \mathbf{y}_g^u, T^{(0)} = \varnothing$;
3: **While** $k < N_a N_{af}$ **do**
4: $k = k+1$, $q = 0$, $T^{(k)} = \varnothing$;
5: **For** $i = 1$ **to** $\left|T^{(k-1)}\right|$ **do**
$[t_1, t_2, ..., t_L] = \arg\text{sort}_{L\ best}\left(\left|\left(\breve{\mathbf{H}}_g^u\right)^H \mathbf{r}_i^{(k-1)}\right|\right)$. (sort $L$ best indices)
6: **For** $j = 1$ **to** $L$ **do**
7: $t_{tmp} = t_i^{(k-1)} \cup \{t_j\}$;
8: **If** ($t_{tmp} \notin T^{(k)}$) **and** (number of indices in $t_{tmp}$ inside subcarrier block of $t_j$ is $\leq N_a$) **and** (number of subcarrier blocks with indices in $t_{tmp}$ is $\leq N_{af}$) **then**
9: $q = q+1$;
10: $t_q^{(k)} = t_{tmp}$;
11: $T^{(k)} = T^{(k)} \cup \{t_q^{(k)}\}$;
12: $\left(\tilde{\mathbf{s}}_g^u\right)^{(k),(q)} \leftarrow 0$
$\left(\tilde{\mathbf{s}}_g^u[t_q^{(k)}]\right)^{(k),(q)} \leftarrow \Pi_{\mathcal{A}^{|t_q^{(k)}|}}\left(\left(\left(\breve{\mathbf{H}}_g^u[:,t_q^{(k)}]\right)^H \breve{\mathbf{H}}_g^u[:,t_q^{(k)}]\right)^{-1}\right.$
$\left.\times \left(\breve{\mathbf{H}}_g^u[:,t_q^{(k)}]\right)^H \mathbf{y}_g^u\right)$
13: $\mathbf{r}_q^{(k)} = \mathbf{y}_g^u - \breve{\mathbf{H}}_g^u[:,t_q^{(k)}]\left(\tilde{\mathbf{s}}_g^u\right)^{(k),(q)}$.
14: **end if**
15: **end for**
16: **end for**
17: **end while**
18: $q^* = \arg\min_q \left\|\mathbf{r}_q^{(N_a N_{af})}\right\|_2$;
19: **Output:** $\hat{\mathbf{s}}_g^u = \left(\tilde{\mathbf{s}}_g^u\right)^{(N_a N_{af}),(q^*)}$.

where $\breve{\mathbf{H}}_g^u = \hat{\mathbf{H}}_g^u\left(\mathbf{A}_{N_f} \otimes \mathbf{I}_{N_s}\right)$ is the overall equivalent channel "seen" by user $u$ for the $g^{\text{th}}$ PT-GSFIM symbol. In the following we propose three different algorithms that can be adopted for the detector. The first one is a generalization of the OB-MMSE from [44] and is shown in Algorithm 1. The main differences in the presented algorithm reside in step 2 and 3 which have to be computed over both the spatial and frequency dimensions and require working with the equivalent channel matrix for the whole PT-GSFIM symbol $\breve{\mathbf{H}}_g^u$. Furthermore, instead of working with the possible antenna combinations, the extended algorithm considers the possible joint combination of active antennas and subcarriers, and sorts these $N_{comb} = 2^{N_{af}\left\lfloor\log_2\binom{N_s}{N_a}\right\rfloor + \left\lfloor\log_2\binom{N_f}{N_{af}}\right\rfloor}$ combinations according to the measured reliability (step 4). In this case we use $\mathbb{I}$ to denote the set of possible supports of $\mathbf{s}_g^u$, i.e., $\mathbb{I} = \{I_1,...,I_{N_{comb}}\}$ with $I_i = \text{supp}\left(\mathbf{s}_g^u\right)$ and $i = 1,2,...,N_{comb}$. Each of these possible antennas and subcarrier combinations is processed using a block MMSE detector applied with a fixed support (step 7). This step involves an element-wise projection over set $\mathcal{A}_0$, denoted as $\Pi_{\mathcal{A}_0^{N_s N_f}}()$. The algorithm runs until the threshold condition is satisfied (step 8), otherwise it will end only after all the combinations are processed in which case the best estimate will be selected.

The second approach is based on the sMMP from [45] which we extend for a multiblock structured signal such as GSFIM. The resulting detector is shown in Algorithm 2. It follows a greedy strategy similar to the orthogonal matching pursuit (OMP) algorithm [46], but instead of performing a tree search along a single path, it uses parallel search to build a list of candidates and selects the best estimate at the end. The algorithm comprises a total of $N_a N_{af}$ iterations. Defining $L$ as the number of child candidates, in each iteration, the algorithm expands each of the previous iteration candidates into $L$ newer candidates with an additional component that is selected so as to maximize the correlation with the previous residual. While the total number of candidates can keep increasing along the iterations, some of them may overlap and can be removed. In order to adapt the algorithm to GSFIM signals, it has to work jointly with the spatial and frequency dimensions and with the equivalent channel matrix for the whole PT-GSFIM symbol.

TABLE III
ALGORITHM 3: ADMM-BASED GSFIM DETECTOR

| | |
|---|---|
| 1: | **Input:** $\mathbf{x}^0, \mathbf{r}^0, \mathbf{z}^0, \mathbf{u}^0, \mathbf{v}^0, \mathbf{w}^0$, $\breve{\mathbf{H}}_g^u$, $\mathbf{y}_g^u$, $\rho_x, \rho_r, \rho_z, Q$ |
| 2: | $f_{best} = \infty$. |
| 3: | $\boldsymbol{\Phi} \leftarrow \left( \left(\breve{\mathbf{H}}_g^u\right)^H \breve{\mathbf{H}}_g^u + (\rho_x + \rho_r + \rho_z)\mathbf{I}_{N_f N_s} \right)^{-1}$. |
| 4: | **for** $t = 0,1,\ldots Q-1$ **do** |
| 5: | $\mathbf{s}^{(t+1)} \leftarrow \boldsymbol{\Phi}\left( \left(\breve{\mathbf{H}}_g^u\right)^H \mathbf{y}_g^u + \rho_x\left(\mathbf{x}^{(t)} - \mathbf{u}^{(t)}\right) + \rho_r\left(\mathbf{r}^{(t)} - \mathbf{v}^{(t)}\right) \right.$ $\left. + \rho_z\left(\mathbf{z}^{(t)} - \mathbf{w}^{(t)}\right)\right)$. |
| 6: | $\left(\mathbf{x}^i\right)^{(t+1)} \leftarrow \Pi_{\mathbb{S}_0}\left(\left(\mathbf{s}^i\right)^{(t+1)} + \left(\mathbf{u}^i\right)^{(t+1)}\right)$. |
| 7: | $\mathbf{r}^{(t+1)} \leftarrow \Pi_{\mathbb{J}}\left(\mathbf{s}^{(t+1)} + \mathbf{v}^{(t)}\right)$. |
| 8: | $\mathbf{z}^{(t+1)} \leftarrow \Pi_{\mathcal{A}_0^{N_s N_f}}\left(\mathbf{s}^{(t+1)} + \mathbf{w}^{(t)}\right)$. |
| 9: | $I \leftarrow \mathrm{supp}\left(\mathbf{x}^{(t+1)}\right) \cap \mathrm{supp}\left(\mathbf{r}^{(t+1)}\right)$. |
| 10: | **If** $t = Q-1$ **then** |
| 11: | $\tilde{\mathbf{s}}_g^u[\bar{I}] \leftarrow 0$ $\tilde{\mathbf{s}}_g^u[I] \leftarrow \Pi_{\mathcal{A}^{N_{af}N_a}}\left( \left( \left(\breve{\mathbf{H}}_g^u[:,I]\right)^H \breve{\mathbf{H}}_g^u[:,I] + 2\sigma^2 \mathbf{I}_{N_{af}N_s} \right)^{-1} \times \right.$ $\left. \times \left(\breve{\mathbf{H}}_g^u[:,I]\right)^H \mathbf{y}_g^u \right)$ (polishing) |
| 12: | **else** |
| 13: | $\tilde{\mathbf{s}}_g^u[\bar{I}] \leftarrow 0$, $\tilde{\mathbf{s}}_g^u[I] \leftarrow \mathbf{z}^{(t+1)}[I]$. |
| 14: | **end if** |
| 15: | **If** $f\left(\tilde{\mathbf{s}}_g^u\right) < f_{best}$ **then** |
| 16: | $\hat{\mathbf{s}}_g^u[\bar{I}] \leftarrow 0$, $\hat{\mathbf{s}}_g^u[I] \leftarrow \tilde{\mathbf{s}}_g^u[I]$. |
| 17: | $f_{best} = f\left(\tilde{\mathbf{s}}_g^u\right)$. |
| 18: | **end if** |
| 19: | $\mathbf{u}^{(t+1)} \leftarrow \mathbf{u}^{(t)} + \mathbf{s}^{(t+1)} - \mathbf{x}^{(t+1)}$. |
| 20: | $\mathbf{v}^{(t+1)} \leftarrow \mathbf{v}^{(t)} + \mathbf{s}^{(t+1)} - \mathbf{r}^{(t+1)}$. |
| 21: | $\mathbf{w}^{(t+1)} \leftarrow \mathbf{w}^{(t)} + \mathbf{s}^{(t+1)} - \mathbf{z}^{(t+1)}$. |
| 22: | **end for** |
| 23: | **Output:** $\hat{\mathbf{s}}_g^u$. |

Furthermore, step 8 has to be modified so that a new candidate is only added to the list if the selected component does not exceed a total of $N_a$ nonzero components in that subcarrier, nor if it results in a total of subcarriers with nonzero components exceeding $N_{af}$.

The third detector is based on the application of ADMM, following an approach that is an extension of the one adopted in [39] for GSM. First, we formulate the maximum likelihood detector (MLD) detection problem as

$$\min_{\mathbf{s}_g^u} \ f(\mathbf{s}) \triangleq \left\| \mathbf{y}_g^u - \breve{\mathbf{H}}_g^u \mathbf{s}_g^u \right\|_2^2 \tag{13}$$

$$\text{subject to} \ \mathbf{s}_g^u \in \mathcal{A}_0^{N_s N_f} \tag{14}$$

$$\mathrm{supp}\left(\mathbf{s}_g^{u,i}\right) \in \mathbb{S}_0, \ i=1,\ldots,N_f \tag{15}$$

$$\mathrm{supp}\left(\mathbf{s}_g^u\right) \in \mathbb{J}, \tag{16}$$

where $\mathcal{A}_0$ represents the complex valued APM constellation set, including symbol 0. $\mathbb{S}_0$ represents the set of possible supports of $\mathbf{s}_g^{u,i}$ according to possible GSM symbols i.e., it is the set with valid (virtual) active antenna combinations including the null support (all antennas inactive). $\mathbb{J}$ denotes the set of possible supports of $\mathbf{s}_g^{u,i}$ according to the valid active subcarrier combinations. It is important to highlight that since this problem formulation is different from the original one presented in [39], mostly due to constraints (15) and (16), the whole algorithm has to be derived again, as will be done next. Introducing auxiliar variables, $\mathbf{x}$, $\mathbf{r}$, $\mathbf{z}$, we can integrate constraints (13)-(14) into the objective function and at the same time make it separable by rewriting the MLD problem as

$$\min_{\mathbf{s}_g^u} \ f\left(\mathbf{s}_g^u\right) \triangleq \left\| \mathbf{y}_g^u - \breve{\mathbf{H}}_g^u \mathbf{s}_g^u \right\|_2^2 + \sum_{i=1}^{N_f} I_{\mathbb{S}_0}(\mathbf{x}^i) + I_{\mathbb{J}}(\mathbf{r}) + I_{\mathcal{A}_0^{N_s N_f}}(\mathbf{z}) \tag{17}$$

$$\text{subject to} \ \mathbf{s}_g^{u,i} = \mathbf{x}^i, \ i=1,\ldots,N_f \tag{18}$$

$$\mathbf{s}_g^u = \mathbf{r} \tag{19}$$

$$\mathbf{s}_g^u = \mathbf{z}. \tag{20}$$

Working with $\mathbf{s}$ instead of $\mathbf{s}_g^u$ in the expressions, the augmented Lagrangian function (ALF) for this problem is given by

$$L_{\rho_x,\rho_r,\rho_z}(\mathbf{s},\mathbf{x},\mathbf{r},\mathbf{z},\mathbf{u},\mathbf{v},\mathbf{w}) = \left\| \mathbf{y}_g^u - \breve{\mathbf{H}}_g^u \mathbf{s} \right\|_2^2 + \sum_{i=1}^{N_f} I_{\mathbb{S}_0}(\mathbf{x}^i) + I_{\mathbb{J}}(\mathbf{r})$$
$$+ I_{\mathcal{A}_0^{N_s N_f}}(\mathbf{z}) + \rho_x\left(\|\mathbf{s}+\mathbf{u}-\mathbf{x}\|_2^2 - \|\mathbf{u}\|_2^2\right) + \rho_r\left(\|\mathbf{s}+\mathbf{v}-\mathbf{r}\|_2^2 - \|\mathbf{v}\|_2^2\right)$$
$$+ \rho_z\left(\|\mathbf{s}+\mathbf{w}-\mathbf{z}\|_2^2 - \|\mathbf{w}\|_2^2\right) \tag{21}$$

where $\rho_x, \rho_r, \rho_z$ are penalty parameters and $\mathbf{u}, \mathbf{v}, \mathbf{w} \in \mathbb{C}^{N_s N_f \times 1}$ are scaled dual variables. We then apply the dual ascent method to the dual problem which requires us to iteratively accomplish the independent minimization of the augmented Lagrangian over $\mathbf{s}$, $\mathbf{x}$, $\mathbf{r}$ and $\mathbf{z}$ combined with the dual variables update through gradient ascent. This results in the following sequence of steps:

• *Step 1: Minimization of the ALF over* $\mathbf{s}$. This step consists in solving

$$\mathbf{s}^{(t+1)} = \min_{\mathbf{s}} \left\{ \left\| \mathbf{y}_g^u - \breve{\mathbf{H}}_g^u \mathbf{s} \right\|_2^2 + \rho_x \left( \left\| \mathbf{s} + \mathbf{u} - \mathbf{x} \right\|_2^2 - \left\| \mathbf{u} \right\|_2^2 \right) \right.$$
$$\left. + \rho_r \left( \left\| \mathbf{s} + \mathbf{v} - \mathbf{r} \right\|_2^2 - \left\| \mathbf{v} \right\|_2^2 \right) + \rho_z \left( \left\| \mathbf{s} + \mathbf{w} - \mathbf{z} \right\|_2^2 - \left\| \mathbf{w} \right\|_2^2 \right) \right\}. \quad (22)$$

A closed-form solution to this problem can be obtained from $\nabla_{\mathbf{s}} L_{\rho_x,\rho_r,\rho_z}(\mathbf{s},\mathbf{x},\mathbf{r},\mathbf{z},\mathbf{u},\mathbf{v},\mathbf{w}) = 0$ which, leads to

$$\mathbf{s}^{(t+1)} = \mathbf{\Phi} \left( \left( \breve{\mathbf{H}}_g^u \right)^H \mathbf{y}_g^u + \rho_x \left( \mathbf{x}^{(t)} - \mathbf{u}^{(t)} \right) + \rho_r \left( \mathbf{r}^{(t)} - \mathbf{v}^{(t)} \right) \right.$$
$$\left. + \rho_z \left( \mathbf{z}^{(t)} - \mathbf{w}^{(t)} \right) \right) \quad (23)$$

with

$$\mathbf{\Phi} = \left( \left( \breve{\mathbf{H}}_g^u \right)^H \breve{\mathbf{H}}_g^u + (\rho_x + \rho_r + \rho_z) \mathbf{I}_{N_f N_s} \right)^{-1}. \quad (24)$$

• *Step 2: Minimization of the ALF over $\mathbf{x}$*. This step solves

$$\mathbf{x}^{(t+1)} = \min_{\mathbf{x}} \left\{ \sum_{i=1}^{N_f} I_{\mathbb{S}_0}(\mathbf{x}^i) + \rho_x \left( \left\| \mathbf{s} + \mathbf{u} - \mathbf{x} \right\|_2^2 - \left\| \mathbf{u} \right\|_2^2 \right) \right\}. \quad (25)$$

The minimization involves an indicator function applied to each individual subcarrier and can be obtained as

$$\left( \mathbf{x}^i \right)^{(t+1)} = \Pi_{\mathbb{S}_0} \left( \left( \mathbf{s}^i \right)^{(t+1)} + \left( \mathbf{u}^i \right)^{(t+1)} \right), \quad i = 1,\ldots,N_f, \quad (26)$$

i.e., as the projection over set $\mathbb{S}_0$, $\Pi_{\mathbb{S}_0}(.)$. This projection can be performed by selecting the $N_a$ largest magnitude elements whose indices also match a valid active antenna combination.

• *Step 3: Minimization of the ALF over $\mathbf{r}$*. The third step solves

$$\mathbf{r}^{(t+1)} = \min_{\mathbf{r}} \left\{ I_{\mathbb{J}}(\mathbf{r}) + \rho_r \left( \left\| \mathbf{s} + \mathbf{v} - \mathbf{r} \right\|_2^2 - \left\| \mathbf{v} \right\|_2^2 \right) \right\}. \quad (27)$$

which also involves an indicator function and whose solution corresponds to the projection over set $\mathbb{J}$, namely

$$\mathbf{r}^{(t+1)} = \Pi_{\mathbb{J}} \left( \mathbf{s}^{(t+1)} + \mathbf{v}^{(t)} \right). \quad (28)$$

This projection can be accomplished by selecting the $N_{af}$ columns of matrix $\mathbf{R} = vec^{-1}_{N_s \times N_f} \left( \mathbf{s}^{(t+1)} + \mathbf{v}^{(t)} \right)$ with larger Euclidean norm and which also match a valid active subcarrier combination, while nulling all the others. The result is then simply set as $\mathbf{r} = vec(\mathbf{R})$.

• *Step 4: Minimization of the ALF over $\mathbf{z}$*. The fourth step solves

$$\mathbf{z}^{(t+1)} = \min_{\mathbf{z}} \left\{ I_{\mathcal{A}_0^{N_s N_f}}(\mathbf{z}) + \rho_z \left( \left\| \mathbf{s} + \mathbf{w} - \mathbf{z} \right\|_2^2 - \left\| \mathbf{w} \right\|_2^2 \right) \right\}. \quad (29)$$

whose solution can also be obtained as a projection, but in this case over set $\mathcal{A}_0^{N_s N_f}$

$$\mathbf{z}^{(t+1)} = \Pi_{\mathcal{A}_0^{N_s N_f}} \left( \mathbf{s}^{(t+1)} + \mathbf{w}^{(t)} \right). \quad (30)$$

It can be implemented elementwise using simple rounding to the closest element in $\mathcal{A}_0$.

• *Step 5: Dual variable update*. The last main step updates the dual variables according to

$$\mathbf{u}^{(t+1)} = \mathbf{u}^{(t)} + \mathbf{s}^{(t+1)} - \mathbf{x}^{(t+1)}, \quad (31)$$
$$\mathbf{v}^{(t+1)} = \mathbf{v}^{(t)} + \mathbf{s}^{(t+1)} - \mathbf{r}^{(t+1)}, \quad (32)$$
$$\mathbf{w}^{(t+1)} = \mathbf{w}^{(t)} + \mathbf{s}^{(t+1)} - \mathbf{z}^{(t+1)}. \quad (33)$$

The overall sequence of steps is summarized in Algorithm 3, where $\bar{I}$ denotes the complement of the support set $I$ (i.e., $\bar{I} = \{1,\ldots,N_s N_f\} \setminus I$) and $Q$ is the maximum number of iterations. It is important to note that the projections of both step 6 and step 7 can be simplified to simple cardinality-based projections if set $\mathbb{S}_0$ or set $\mathbb{J}$ are very large, i.e., there is a big number of valid active antenna combinations and/or valid active subcarrier combinations. In this case, projection $\Pi_{\mathbb{S}_0}()$ is obtained by zeroing the $N_s - N_a$ smallest magnitude elements whereas $\Pi_{\mathbb{J}}()$ is computed by zeroing the $N_f - N_{af}$ columns of $\mathbf{R}$ with smaller Euclidean norm. Algorithm 3 includes a final polishing procedure (step 11) where the projected MMSE estimate is computed for the reduced problem with the support fixed. Regarding the initialization of the algorithm, the warm and random start procedures described in [39] can be directly extended. Note that compared with [39], the proposed ADMM algorithm works with the joint spatial and frequency dimensions and besides several differences in previously existing steps, it also includes two main additional steps, namely: an additional projection (step 7) and an additional dual variable update (step 20).

## B. Computational Complexities

In Table IV, we present the respective worst-case complexities in number of real floating-point operations (flops) of the three proposed algorithms alongside the maximum likelihood detector. In the case of GSFIM-OB-MMSE, $N_{comb}$ corresponds to the number of possible supports of $\mathbf{s}_g^u$ and is given by

$$N_{comb} = 2^{N_{af} \left\lfloor \log_2 \binom{N_s}{N_a} \right\rfloor + \left\lfloor \log_2 \binom{N_f}{N_{af}} \right\rfloor}. \quad (34)$$

In the presented complexities it is assumed that no earlier termination of the GSFIM-ADMM algorithm occurs. Furthermore, to arrive at the expressions we considered that each complex-valued sum, multiplication and squared absolute value computation counts as 2, 6 and 3 real flops. We also took into account that some matrix operations do not need to be repeated after the first iteration (ex: matrix multiplication $\left(\breve{\mathbf{H}}_g^u\right)^H \mathbf{y}_g^u$ in step 5 of GSFIM-ADMM algorithm). In section V we use the presented complexity expressions to compare the three algorithms in different scenarios.

TABLE IV
NUMBER OF REAL FLOPS FOR THE DIFFERENT GSFIM DETECTORS

| Detector | Complexity |
|---|---|
| MLD | $\left( 8N_{rx}N_f N_a N_{af} + 4N_f N_{rx} - 1 \right) N_{comb} M^{N_a N_{af}}$ |
| GSFIM-OB-MMSE | $12 N_{rx} N_f N_a N_{af} + 2 N_s N_f$ $+ N_{comb} \left( N_a N_{af} - 1 + 4(N_a N_{af})^3 + 12(N_a N_{af})^2 N_{rx} N_f + \right.$ $\left. + 7(N_a N_{af})^2 + 14 N_a N_{af} N_{rx} N_f + 4 N_{rx} N_f - 1 \right)$ |
| GSFIM-sMMP ($L=1$) | $\left( 8 N_s N_f^2 N_{rx} + N_s N_f \right) N_a N_{af} + \left( 5 N_{rx} N_f - 2 \right)$ $+ \sum_{k=1}^{N_a} \left( 4k^3 + k^2 (4 N_{rx} N_f + 15) + k(20 N_{rx} N_f - 5) \right)$ |
| GSFIM-sMMP ($L>1$) | $\left( 8 N_s N_f^2 N_{rx} + N_s N_f \right) (1 - L^{N_a N_{af}})/(1-L) + (5 N_{rx} N_f - 2) L^{N_a N_{af}}$ $+ \sum_{k=1}^{N_a} \left( 4k^3 + k^2 (4 N_{rx} N_f + 15) + k(20 N_{rx} N_f - 5) \right) L^k$ |
| GSFIM-ADMM | $4(N_s N_f)^3 + (N_s N_f)^2 (4 N_{rx} N_f + 7) + N_s N_f (12 N_{rx} N_f - 1)$ $+ Q \left( 8(N_s N_f)^2 + 59 N_s N_f - 6 N_f \right) + 4(N_a N_{af})^3$ $+ (N_a N_{af})^2 (4 N_{rx} N_f + 15) + N_a N_{af} (12 N_{rx} N_f - 3)$ |

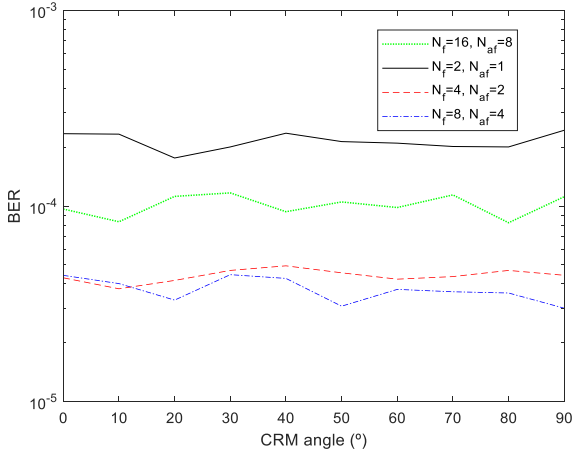

**Fig 3.** Comparison of CRM rotation angle φ of PT-GSFIM in a MU scenario with $N_u$=4, $N_{tx}$=8$N_u$, $N_f$=4, $N_{af}$=3, $N_s$=8, $N_a$=1, $N_{rx}$=4, QPSK (SNR=0dB).

## V. PERFORMANCE RESULTS

In this section, the performance of the proposed PT-GSFIM scheme with the different detection algorithms are assessed and compared against other MU-MIMO systems. Monte Carlo simulations were run according to the system model presented previously. The adopted channel model was the Extended Typical Urban model (ETU) [47] (similar conclusions could be reached for other severely time-dispersive channels). It is assumed that all the channel coefficients are independently drawn according to a complex Gaussian distribution $\mathcal{CN}(0,1)$, and that all users experience the same path-loss. The symbols $s_{g,f}^{u,i}$ transmitted in the active positions (subcarrier and antenna) are randomly selected from an $M$-QAM constellation with equal probabilities and with $E\left[\left|s_{g,f}^{u,i}\right|^2\right]=1$. To understand the impact of the CRM rotation angle φ on the performance of PT-GSFIM, Fig. 3 shows the BER results obtained for different sizes of $N_f$. We assume a scenario with $N_u$=4, $N_{tx}$=8$N_u$, $N_s$=8, $N_a$=1, $N_{rx}$=4, SNR=0 dB (per user), QPSK and GSFIM-ADMM detector. It can be observed that in general the results do not exhibit significant sensitivity to the rotation angle (variations are due to simulation accuracy). What is relevant is the spreading of the PT-GSFIM symbols over several subcarriers through the use of CRM combined with the interleaver in order to obtain additional diversity. It is important to note that even though this figure can show better BERs for smaller $N_f$, larger $N_f$ tend to achieve better results at higher signal-to-noise ratios (SNRs) as we will show further ahead.

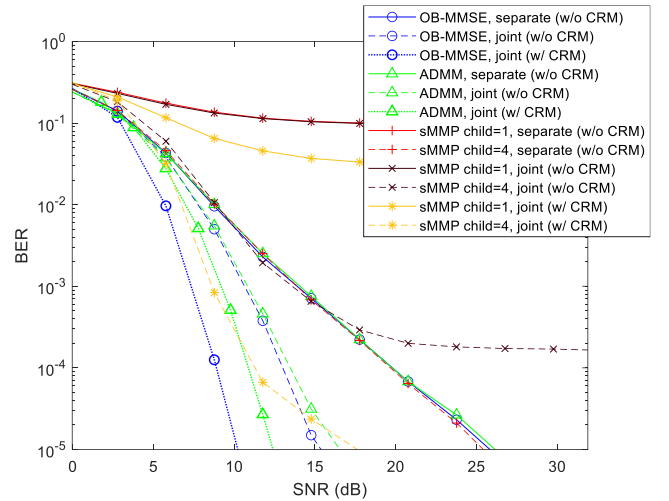

**Fig 4.** BER performance of PT-GSFIM in a MU scenario with $N_u$=4, $N_{tx}$=5$N_u$, $N_f$=4, $N_{af}$=3, $N_s$=5, $N_a$=2, $N_{rx}$=5, QPSK (5.75 bpcu per user).

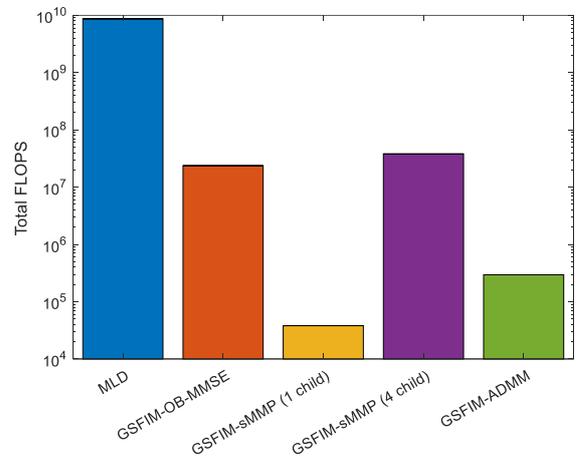

**Fig 5.** Complexity of the different GSFIM detector algorithms for the MU scenario with $N_u$=4, $N_{tx}$=5$N_u$, $N_f$=4, $N_{af}$=3, $N_s$=5, $N_a$=2, $N_{rx}$=5, QPSK (5.75 bpcu per user).

Next, we evaluate the performance of the different receivers and the impact of the use of CRM in PT-GSIFM, with a rotation angle φ of 30º and reminding that the CRM is applied to each set of $N_f$ subcarriers. We assume a base scenario with $N_u$=4, $N_{tx}$=5$N_u$, $N_f$=4, $N_{af}$=3, $N_s$=5, $N_a$=2 and $N_{rx}$=5, which corresponds to a SE of 5.75 bits per channel use (bpcu) and per user. In Fig. 4, three different BER curves are shown for each of the proposed detectors. The "separate" curve corresponds to the direct application of the GSM detector preceded by an active subcarrier detection step. In this case no CRM is applied. The curves where "joint" appears in the legend consider the joint detection of the GSFIM symbols using Algorithm 1-3. For this case, curves with and without CRM are included. It can be observed that the joint detection clearly provides substantial gains over the "separate" approach. The OB-MMSE algorithm achieved the best results followed by ADMM (with 10% of penalty). As for sMMP, it can be seen that it clearly underperforms. It is important to note that sMMP relies on consecutively finding, through correlation, the columns of the channel matrix that are more closely related the residuals of the constructed candidates (step 5 of algorithm 2). Since the cross-correlation between the different columns of the channel matrix has no guarantee of being close to 0 for the scenario of Fig. 4 (channel matrix is random and squared), this will tend to result in errors even if no noise exists, causing irreducible BER floors. Increasing the number of child nodes, $L$, combined with CRM can improve the results, lowering the error floor. In fact, in general, the use of CRM has a large positive impact on the performance of the studied receivers. For example, when using OB-MMSE, CRM provides a gain of 5 dB at a BER of $10^{-5}$. It is important to note that even though OB-MMSE with CRM gives the best results with QPSK, it requires a higher computational complexity than the other approaches, as $N_f$ increases. This can be observed in Fig. 5, where we plot the complexity in flops of the different algorithms, using the expressions provided in section IV.B. As reference we also show the complexity of the MLD whose total number of flops is several orders of magnitude higher than the other alternatives, making it an unpractical solution. We can also see that the lowest complexity approach is the GSFIM-sMMP with 1 child node. However, this case revealed a poor performance in Fig. 4. Increasing the number of child nodes improves the BER results of GSFIM-sMMP but also increases its complexity, making GSFIM-ADMM the approach with the best tradeoff in terms of performance and complexity.

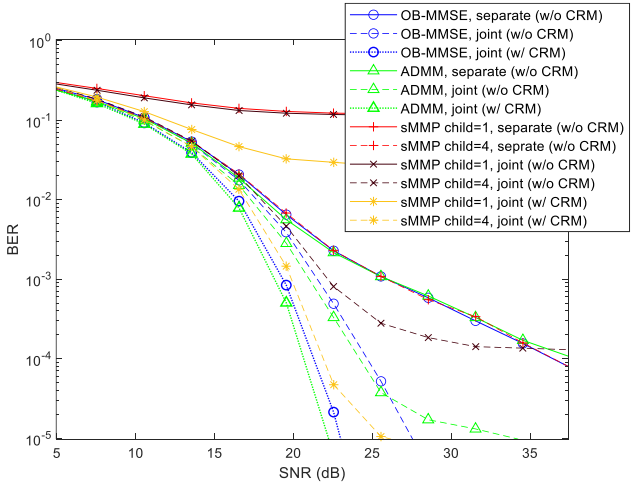

Fig 6. BER performance of PT-GSFIM in a MU scenario with $N_u=4$, $N_{tx}=5N_u$, $N_f=4$, $N_{af}=3$, $N_s=5$, $N_a=2$, $N_{rx}=5$, 64-QAM (11.75 bpcu per user).

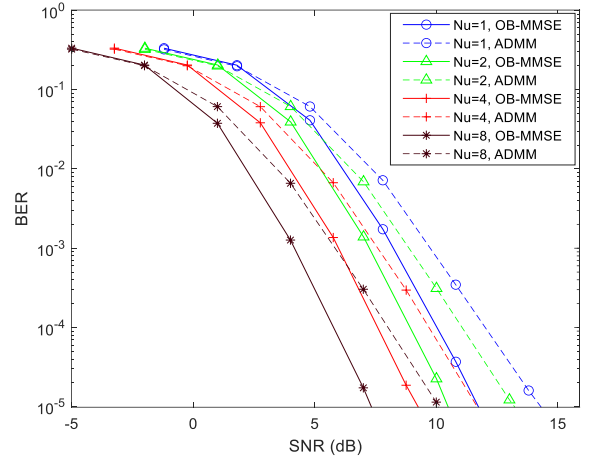

Fig 7. BER performance of PT-GSFIM in a MU scenario with $N_{tx}=5N_u$, $N_f=4$, $N_{af}=3$, $N_s=5$, $N_a=2$, $N_{rx}=4$, QPSK (5.75 bpcu per user).

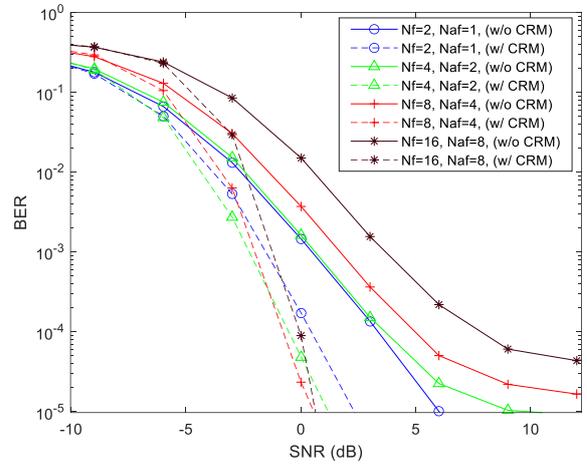

Fig 8. BER performance of PT-GSFIM in a MU scenario with $N_u=4$, $N_{tx}=8N_u$, $N_s=8$, $N_a=1$, $N_{rx}=4$, QPSK (3 bpcu per user).

Based on the previous scenario, the signal constellation was increased from QPSK to 64-QAM. The results are shown in Fig. 6. Most of the relative behavior of the curves observed in Fig. 4 repeats itself in this graph. However, in this case, ADMM achieves the best results followed by OB-MMSE. As verified previously, the use of CRM clearly improves the results.

Next, we study the influence of the number users on the performance of the receivers, considering basically the same scenario of Fig. 4, with a QPSK signal constellation and the use of CRM. Fig. 7 shows that independently of the chosen algorithm, the results improve by increasing the number of users. The improvement is achieved due to the additional transmit antennas (since $N_{tx}=5N_u$) combined with the power minimization step applied at the transmitter (detailed in [34]). It is important to note also that due to the use of block diagonalization based precoders, which remove all multiuser interference and transform the signal arriving at each user to an equivalent SU channel (as shown in eq. (8)), there is no loss of diversity in this scenario. In fact, since we always use $N_{tx}=5N_u$ and $N_s=5$ when we increase the number of users, the equivalent received SU model (eq. (8)) always maintains the same dimensions.

In the next scenario, which is shown in Fig. 8, we evaluate the impact of the CRM size, defined according to different values of $N_f$ and $N_{af}$. Due to the very high computational complexity of OB-MMSE for large CRMs, we consider the adoption of the ADMM algorithm only. The system is operating with $N_u=4$, $N_{tx}=8N_u$, $N_s=8$, $N_a=1$, $N_{rx}=4$, and QPSK. Note that the SE of the different curves is not always exactly the same, but it stays very close to 3 bpcu per user (varies between 3 and 3.3). As reference, curves without CRM are also included. It can be observed that without CRM, using larger subcarrier subblocks ($N_f$) tends to degrade the performance due to the more challenging detection task making the curves exhibit irreducible BER floors. In fact, it is important to note this figure corresponds to a difficult underdetermined scenario, where the number of observations is smaller than the number of unknowns. With CRM, the behavior is reversed. Even though, for low SNRs a larger $N_f$ achieves worse BERs, at higher SNRs the curves fall with a steeper drop and end up achieving better results. Comparing the curves of PT-GSFIM with CRM against the curves without CRM for the same dimensions in term of $N_f$ and $N_{af}$, we can observe gains between

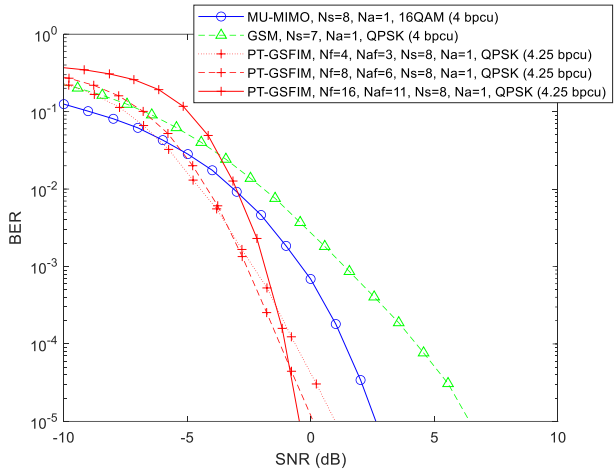

**Fig 9. BER performance of PT-GSFIM, PT-SDIM, PTSFIM and conventional BD MU-MIMO (Nu=8).**

3 dB (smallest $N_f$) and 8 dB (largest $N_f$) for a BER of $10^{-4}$, due to the additional diversity.

In the last set of results shown in Fig. 9, we present a comparison between multiple scenarios using a conventional Block Diagonalization (BD) MU-MIMO scheme from [43], GSM MU-MIMO from [34], and the proposed PT-GSFIM. Regarding PT-GSFIM, we present three different configurations, while keeping the same SE. For all the curves we consider $N_u$=8, with the SE being close to 4 bpcu per user. ADMM based detection was adopted for all the schemes, i.e., for BD MU-MIMO, GSM MU-MIMO and PT-GSFIM. It can be observed that the proposed PT-GSFIM schemes can outperform both GSM and conventional MU-MIMO and the gains are higher when the symbols are spread over a larger number of subcarriers ($N_f$).

## VI. CONCLUSION

In this paper we described a MU-MIMO system where a base station transmits precoded space-frequency domain IM symbols. The proposed PT-GSFIM scheme adopts SSD techniques which allows it to benefit from the diversity effects inherent to a frequency selective channel. To support the detection of PT-GSFIM signals at the receiver, three different algorithms were presented. As expected, it was observed that by increasing the size of the subcarrier subblocks while keeping the same SE, can result in improved performance when SSD is employed, since the symbols are spread over a larger number of subcarriers, benefiting from the frequency diversity effect. Performance results also demonstrated that the proposed scheme can outperform both GSM and conventional MU-MIMO.

### ACKNOWLEDGMENT

This work was partially supported by the FCT - Fundação para a Ciência e Tecnologia under the grant 2020.05621.BD. The authors also acknowledge the funding provided by FCT/MCTES through national funds and when applicable co-funded EU funds under the project UIDB/50008/2020.